\documentclass[pre,aps,twocolumn,superscriptaddress,showpacs]{revtex4-1}
\usepackage{bm}
\usepackage{amssymb}
\usepackage{amsfonts}
\usepackage{color}
\usepackage[dvips]{graphicx}
\usepackage[cmex10]{amsmath}
\usepackage{algorithmic}

\newcommand{\beq}{\begin{equation}}
\newcommand{\eeq}{\end{equation}}
\newcommand{\bea}{\begin{eqnarray}}
\newcommand{\eea}{\end{eqnarray}}

\newcommand{\rv}{\underline{\bf{r}}}
\newcommand{\Fv}{\underline{\bf{F}}}
\newcommand{\sv}{\underline{\bf{S}}}

\newcommand{\nv}{\underline{\bf{n}}}

\newcommand{\vect}[1]{\underline{\mathbf{#1}}}

\newtheorem{algorithm}{Algorithm}
\DeclareMathOperator*{\argmax}{arg\,max}
\DeclareMathOperator*{\argmin}{arg\,min}

\begin{document}

\title{Information Storage and Retrieval for Probe Storage using Optical Diffraction Patterns}
\date{\today}

\author{\fbox{Joost~W.~van~Honschoten}}
\affiliation{MESA$+$ Research
Institute, University of Twente, P.O. Box $217$, $7500$ AE, Enschede
The Netherlands}
\author{Henri~W.~de~Jong}
\email{henridejong@gmail.com}
\affiliation{MESA$+$ Research
Institute, University of Twente, P.O. Box $217$, $7500$ AE, Enschede
The Netherlands}
\author{Wabe~W.~Koelmans}
\email{W.W.Koelmans@alumnus.utwente.nl}
\affiliation{MESA$+$ Research
Institute, University of Twente, P.O. Box $217$, $7500$ AE, Enschede
The Netherlands}
\author{Thomas~P.~Parnell}
\email{tom.parnell@siglead.com}
\affiliation{Siglead Europe Ltd., International Digital Laboratory, University of Warwick, Gibbet Hill Road, Coventry CV4 7AL, UK}
\author{Oleg Zaboronski}
\email{O.V.Zaboronski@warwick.ac.uk}
\affiliation{Mathematics Institute, University of Warwick, Gibbet Hill Road, Coventry CV4 7AL, UK}

\begin{abstract}
A method for fast information retrieval from a probe storage device is considered.
It is shown that information can be stored and retrieved using the optical
diffraction patterns obtained by the illumination of a large array of cantilevers by a monochromatic
light source.
In thermo-mechanical probe storage, the information is stored as a sequence of
indentations on the polymer medium.
To retrieve the information, the array of probes is actuated by applying
a bending force to the cantilevers. Probes positioned over indentations experience deflection by the depth
of the indentation, probes over the flat media remain un-deflected. Thus the array of actuated probes
can be viewed as an irregular
optical grating, which creates a data-dependent diffraction pattern when illuminated by laser light.
We develop a low complexity modulation scheme, which allows the extraction of information stored in the pattern
of indentations on the media from Fourier coefficients of the intensity of the diffraction pattern.
We then derive a low-complexity maximum-likelihood sequence detection algorithm for retrieving
the user information from the Fourier coefficients. The derivation of both the modulation and the detection
schemes is based on the Fraunhofer formula for data-dependent diffraction patterns.
The applicability of Fraunhofer diffraction theory to the optical set-up relevant for probe storage
is established both theoretically and experimentally.
We confirm the potential of the optical readout technique by demonstrating that the impairment
characteristics of probe storage channels (channel noise, global positioning errors, small indentation depth)
do not lead to an unacceptable increase in data recovery error rates.
We also show that for as long as the Fresnel number $F\leq 0.1$, the optimal channel detector derived
from Fraunhofer diffraction theory does not suffer any significant performance degradation.
\end{abstract}


\maketitle

\section{Introduction}

Probe based data storage with the information stored thermo-mechanically as indentations
in a polymer medium is demonstrated in \cite{Pantazi2008}.  Information is written by
heating the cantilever tip and applying a force to result in an indentation being created in
the medium.  Information retrieval is achieved thermally: the resistance of an integrated
resistive heater is monitored as the tip scans the surface. This resistance drops naturally
with cantilever/medium separation due to the decrease in heat transfer from the medium to the heater.
In order to achieve the data rates required by modern storage applications, a probe storage
system will typically consist of a large array of cantilevers reading and writing information in parallel.

In this paper we consider the retrieval of information using the optical
diffraction patterns resulting from illumination of the {\it linear} cantilever array with a laser light source.
The main idea is that different combinations of deflected and un-deflected cantilevers in the array
result in different diffraction patterns, which can therefore be used to store information.
Such an approach to read-back is potentially interesting since
it is fast (the intensity of light can be captured almost instantaneously with a photodiode)
compared to the thermo-resistive alternative (where we must wait for the thermal sensor to reach equilibrium temperature).
One of the disadvantages of the proposed solution is the increased size of the storage
device as the optical path to detect cantilever deflection
will occupy more volume than integrated sensors. Therefore the method of information retrieval
described in our paper targets larger form factors of probe storage, for example back-up and archival storage.

Sensing of cantilever deflection by using optical diffraction patterns has been
introduced by Manalis et al.~\cite{Manalis1996,Yaralioglu1998}
By fabricating a cantilever containing interdigitated fingers a diffraction
grating is formed. When the cantilever is bent, alternating fingers are deflected while
remaining fingers are held fixed. The work is extended to parallel operation by illuminating
multiple interdigital cantilevers and capturing the diffraction pattern from each
cantilever on a separate detector~\cite{Sulchek2001}. Others have fabricated a similar
design where the fixed fingers are attached to a frame surrounding the cantilever and
the authors recognized the potential of sensing in wave-vector space \cite{Thundat2000}.
Recently it has been shown that the intensity of the principle maximum of a diffraction
pattern can be used to detect cantilever deflections \cite{Koelmans2010}. In this case
different natural resonance frequencies are used to address each cantilever.
In \cite{Sache2007} surface topography has been imaged by using a digital holographic microscope
to capture amplitude and phase
information of the light reflected from an array of cantilevers illuminated with a laser source.

We present a scheme where the cantilevers themselves form an
optical grating and the state of the array of cantilevers
can be retrieved by analyzing the full diffraction pattern, which is captured by an array of detectors.

In order to retrieve the state of deflection of the cantilevers using low complexity
signal processing algorithms a special modulation scheme for storing information
inside diffraction patterns is developed. Binary user information is converted to
ternary \textit{trits} which are then mapped to a sequence
of indentations to be recorded on the medium.
We show that by analysing the resulting diffraction patterns
in the Fourier domain it is possible to determine the
most-likely sequence of trits in linear time using a
fast recursive graph-based detection algorithm that
resembles the Smith-Waterman algorithm used for genetic sequence matching \cite{Smith1981}.

The paper is organised as follows. In Section \ref{sec:chan_model} the
channel model is derived using Fraunhofer diffraction theory. The principle of information
storage inside diffraction patterns using the \textit{central trits} is explained.
The channel model is then extended to capture the effects of electronics noise
in the system, the signal-to-noise ratio is defined which allows
us to determine the optimal indentation depth. The model is then compared
with experimental diffraction patterns from an array of cantilevers
and a high level of agreement is observed. Two detection algorithms are presented in
Section \ref{sec:detect}: a simple threshold detector and the optimal maximum-likelihood
sequence detector and their performance is compared. In Section \ref{sec:further} we consider
the effect of the global positioning errors that characterise highly-parallel probe storage.
It is shown that using the techniques introduced in \cite{Parnell2010}
it is possible to construct an asymptotically optimal iterative sequence detector for a large array of probes.
We then proceed to study the effects of sub-optimal pit-depth and modeling noise on the
error-rate of the optical data retrieval algorithms. In Section \ref{sec:conclusions}
the main results of the paper are summarised.
In the appendices a derivation of the channel model using Helmholtz-Kirchhoff diffraction theory is presented.

\section{Channel Model}\label{sec:chan_model}
\begin{figure}
\begin{centering}
\includegraphics[width=9.0cm]{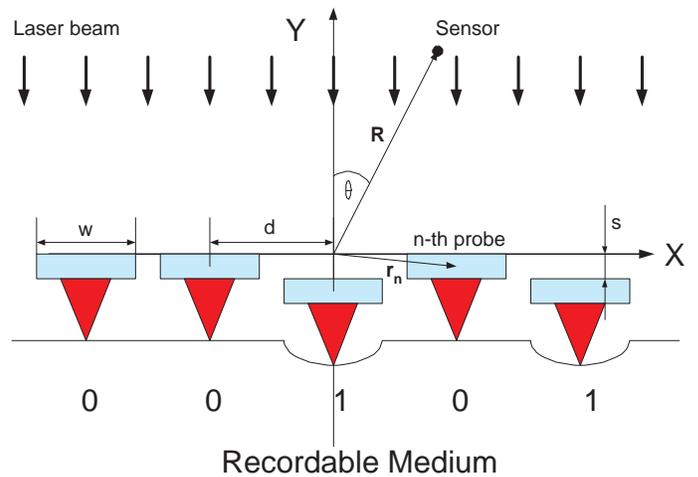}
\par
\end{centering}
\caption{Geometry of the optical readout.}
\label{fig:geom_array}
\end{figure}

\subsection{Storing Information In Diffraction Patterns}

Consider a linear array of $N$ reflective cantilevers with sensing probes positioned above an area of
the medium to which binary information string $b_0~ b_1~\ldots b_{N-1}$ has
been recorded using the thermo-mechanical write process (see Figure \ref{fig:geom_array}). The value $b_n=1$
corresponds to an indentation in the medium, $b_n=0$ - to the absence of indentation, see \cite{Marcellino}
for a review of thermo-mechanical probe storage. During the process of reading, individual probes
are pressed against the polymer surface. As a result,
the $n$-th cantilever can be in two states: deflected if $b_n=1$ or not deflected if $b_n=0$.
Therefore the set of $N$ reflective cantilevers can be viewed as an irregular optical grating, the irregularity being determined
by the data string recorded on the medium.
If the reflective cantilever array is illuminated by a laser source of wavelength $\lambda$ then
a data-dependent diffraction pattern is created. In the practically relevant case of very long cantilevers,
the diffraction pattern is two-dimensional - the distribution of the reflected electromagnetic field depends on
the diffraction angle $\theta$ and distance $R$ to the observation point only, see Figure \ref{fig:geom_array}.

The fact that different binary strings lead to different diffraction patterns allows one to store information
directly in the patterns rather than in individual indentations. The simplest map from binary data to
diffraction patterns would be an enumeration of all distinct patterns.

The optical sensors we used in our experiments can only detect the field's intensity, but not the phase.
The principal fundamental question we need to answer is: how many bits of information can be stored in the intensity of diffraction
patterns resulting from $2^N$ possible configurations of $N$ cantilevers?

All the length scales characterising the array are much greater than wavelength $\lambda$,
see Section \ref{subsec:exper} for details of the experimental set-up. Therefore, we
can use Fraunhofer diffraction theory to calculate the distribution of intensity of reflected light,
see \cite{Born} for a review.

In the Fraunhofer limit any component of the reflected electromagnetic field $U$ can be written as a
function of angle parameter $q=k\theta$, where $k=2\pi/\lambda$ is the wave number:
\begin{eqnarray}
U(q)=C(q)\sum_{n=0}^{N-1}\xi(n)e^{-iqnd}\label{eq:diff_patt}
\end{eqnarray}
where $d$ is the cantilevers' pitch (see Figure \ref{fig:geom_array}) and
\begin{eqnarray}
\xi(n)=e^{2iksb_{n}}
\end{eqnarray}
is the additional phase gained at the $n$-th cantilever deflected by the distance $b_n s$,
$b_{n}\in\{0,1\}$ is the $n$-th bit of the information sequence; $C(q)$ is a \emph{data-independent}
complex amplitude,
\bea
C(q)=A_0\sqrt{\frac{kw^2}{2\pi R}}\frac{\sin(qw/2)}{(qw/2)},
\eea
where $R$ is the distance to the observation point, $w$ is the cantilever's width, the amplitude
$A_0$ is related to the intensity of the laser source and reflective properties of cantilevers.

For the sake of completeness, we placed the derivation of formula (\ref{eq:diff_patt}) along with
a careful analysis of applicability conditions in the appendices.

The intensity $I(q)=\overline{U(q)}{U(q)}$ of the reflected light can be written as follows:
\begin{eqnarray}
I(q)=|C(q)|^2\sum_{n=-(N-1)}^{N-1}f(n)e^{-iqnd}\label{eq:intensity}
\end{eqnarray}
with coefficients $f(n)$ given by:
\begin{eqnarray}
f(n)=\left\{\begin{array}{ll}
\sum_{p=0}^{N-1-n}e^{2ikst_{n+p,p}} & \mbox{ if $n \geq 0$}\\\\
\overline{f(-n)} & \mbox{ if $n < 0$ }\end{array}\right.
\label{eq:fcoeff}
\end{eqnarray}
where $$t_{n+p,p}=b_{n+p}-b_{p}\in\{-1,0,+1\}.$$
The variables $t_{n+p,p}$'s can be viewed as digits of a balanced ternary representation of numbers and
will be referred to as \emph{trits}.

Using (\ref{eq:intensity}), it is very easy to check that the following two information strings lead to
the same intensity pattern $I(q)$:
\begin{eqnarray*}
S_1&=&b_0 b_1\ldots b_{N-2} b_{N-1}\\
S_2&=&\bar{b}_{N-1} \bar{b}_{N-2} \ldots  \bar{b}_1 \bar{b}_0,
\end{eqnarray*}
where $\bar{b}=1-b$ is the operation of bit inversion.
In other words the transformation of
$$T_N: b_n\rightarrow \bar{b}_{N-1-n},~n=0,1,\ldots N-1$$ does not change
the intensity pattern. An exhaustive numerical check for $N=1,2,\ldots, 10$ shows that there is
no other transformation of the information string which leaves the intensity pattern invariant.

Therefore, we can compute the number of distinct diffraction patterns corresponding to $2^N$ binary information
strings as the number of orbits of the transformation $T_N$. If $N$ is odd, all orbits have size $2$. Therefore
the number of distinct diffraction patterns is $2^{N-1}$. If $N$ is even, the calculation is slightly
more difficult as $T_N$ has $2^{N/2}$ fixed points (strings which are invariant under $T_N$) as well as
orbits of size $2$. The final answer is
\begin{eqnarray*}
\sharp_{\mbox{distinct patterns}}(N)=\left\{\begin{array}{ll}
2^{N-1} & \mbox{ if $N$ is odd,}\\\\
2^{N-1}+2^{N/2-1} & \mbox{ if $N$ is even.}\end{array}\right.
\end{eqnarray*}
We conclude that a diffraction pattern created by an $N$-bit binary strings recorded
on the medium contains at least $log_2 2^{(N-1)}=N-1$ bits of information.

How can we retrieve this information? The most straightforward way would be to enumerate
all patterns using $(N-1)$-bit strings. The optimal receiver would then compare the pattern
generated by the sensors with all $2^{N-1}$ model patterns stored in its memory and output
the index of the model pattern which is the most probable transmitted pattern given the received one.
This solution is clearly impractical, as the complexity of the optimal receiver grows exponentially
with $N$ - the number of probes in the array. The rest of the paper is dedicated to the design
of modulation and detection schemes which would allow low complexity storage and retrieval of information
using diffraction patterns. The greatest difficulty we must overcome stems from the non-linear relationship
between the information bits and the corresponding intensity pattern.

\subsection{Ternary Modulation Scheme}
The expression (\ref{eq:intensity}) for the intensity of the diffraction pattern is a product of data-independent
function $|C(q)|^2$ and a band-limited function $\sum_{n=-N+1}^{N-1} f(n) e^{-i q n d}$. Therefore, to extract
the information-dependent coefficients $\{f(n)\}$ from $I(q)$ we need to sample it at finitely many points.
Given that there is no resonant relationship between the cantilevers' width $w$ and the array's period $d$,
these sampling points can be chosen not to coincide with zeros of $|C(q)|^2$. The sequence of sample values
$I(q_1), I(q_2), \ldots$  can be then normalized by $1/|C(q_1)|^2, 1/|C(q_2|^2, \ldots$ using for example a variable
gain amplifier. This remark allows us to assume from now on that $|C(q)|^2=1$ and represent the normalized
intensity of reflected light as
\bea
I(q)=\sum_{n=-N+1}^{N-1} f(n) e^{-i q n d}.\label{eq:intensity1}
\eea

Sampling (\ref{eq:intensity1}) at $2N-1$ uniformly spaced sampling points
$$q_m=\frac{2\pi m}{(2N-1)d},$$
where $m=-(N-1), -(N-2), \ldots, -1, 0, 1, \ldots, N-2, N-1$,
we get a system of $2N-1$ linear equations for $2N-1$ information-carrying coefficients $f(n)$'s.
This system can be solved using a discrete Fourier transform (DFT, \cite{Rivest}):

\begin{eqnarray}
f(n) = \frac{1}{2N-1}\sum_{m=-N+1}^{N-1} I(q_m)e^{inq_md}
\end{eqnarray}
Note that the sampling scheme chosen above is as economical as possible according to
the Nyquist-Shannon sampling theorem \cite{Nyquist}: the linear sampling
frequency is equal to $\frac{(2N-1)d}{2\pi}$ - just over twice the maximal
linear frequency contributing to (\ref{eq:intensity1}):
$\nu_{max} = \frac{(N-1)d}{2\pi}$.

We conclude that the problem of extracting the data-dependent coefficients
$\{f(n)\}$ from the intensity pattern $I(q)$ can be solved at the cost $O(NlogN)$
using the Discrete Fast Fourier Transform algorithm (DFFT).

We are now ready to describe a lossy modulation-demodulation scheme which would allow us
to perform the operations \emph{Information}$\rightarrow$\emph{Bit pattern written on storage medium}
$\rightarrow I(q)\rightarrow$ \emph{Information} at a cost which scales only polynomially with $N$.

We firstly note that since the trit $t_{p,q}\in \{-1,0,1\}$, $t_{p,q}^{2n}=t_{p,q}^2$
and $t_{p,q}^{2n+1}=t_{p,q}$ for any integer $n$. This simple remark allows us to compute
 the real and imaginary parts of $f(n)$:
\begin{eqnarray}
\Re\left(f(n)\right)&=&\sum_{p=0}^{N-1-n}\left(1+(\cos(2ks)-1)t_{n+p,p}^{2}\right)\label{eq:fnr}\\
\Im\left(f(n)\right)&=&\sum_{p=0}^{N-1-n}\sin(2ks)t_{n+p,p}\label{eq:fni}
\end{eqnarray}
While the real part is a non-linear function of the trits the
imaginary part is linear and crucially it can be written as
the sum of the first $n$ \emph{central trits}
\bea
t_{N-1-p,p}=b_{N-1-p}-b_p, ~0\leq p\leq \frac{N-1}{2}.
\eea
Namely, we have the following elementary observation:
\begin{eqnarray}
\Im\left(f(n)\right)&=&\sin(2ks)\sum_{p=0}^{n-1}t_{N-1-p,p}\label{thm}
\end{eqnarray}

We propose to store information in central trits since they can be easily recovered from the output of the DFFT
using equation (\ref{thm}):
\begin{eqnarray}
t_{N-1-n,n} = \frac{\Im\left(f(n+1)\right)}{\sin(2ks)} - \frac{\Im\left(f(n)\right)}{\sin(2ks)}\label{eq:trit_recovery_no_noise}
\end{eqnarray}

In what follows we will often use the shorthand notation for central trits: $t_p=t_{N-1-p,p}$.

\begin{figure}
\begin{centering}
\includegraphics[width=7.0cm]{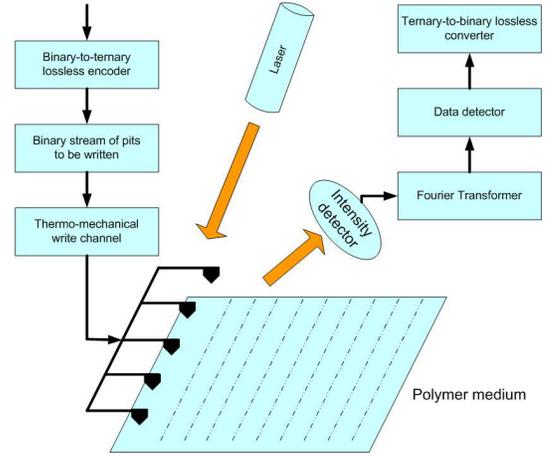}
\par\end{centering}
\caption{Optical Channel Flow}
\label{fig:optical_flow}
\end{figure}

The method of storing and retrieving information in diffraction patterns is now
described with assistance from Figure \ref{fig:optical_flow}.
Firstly the binary user information can be converted to balanced
ternary using a lossless encoder, see e. g. \cite{Knuth}.
The ternary sequence $\vect{t}=(t_{N-1,0},t_{N-2,1},\ldots,t_{N/2-1,N/2})$
is then mapped to the binary sequence $\vect{b}$ to be recorded on the medium as follows: for $n=0,\ldots,N/2-1$:
\begin{eqnarray}
(b_{N-1-n},b_n) = \left\{\begin{array}{ll}
( 0, 1 ) & \mbox {if $t_{N-1-n,n}=-1$} \\
( 0, 0 ) & \mbox {if $t_{N-1-n,n}=0$} \\
( 1, 0 ) & \mbox {if $t_{N-1-n,n}=+1$} \end{array}\right.
\end{eqnarray}
(Note the above map between ternary information string
and binary indentation sequence leads to the loss of information rate: to store information
we use only three out of possible four configurations of two cantilevers.)
At the receiving end,
the intensity of the diffraction pattern can then be measured
and the coefficients $f(n)$ recovered using DFT.
The remaining task is to determine the sequence $\vect{t}$ and convert back to binary to return to the user.
In the absence of noise the central trits are simply given by (\ref{eq:trit_recovery_no_noise}).
In the presence of noise we must \textit{detect} the sequence of trits - this will be discussed in the next section.
The final step is to convert the ternary sequence back to binary.

A close examination reveals that the computational cost of our modulation scheme is bounded by $O(Nlog(N))$,
the most expensive step being the Discrete Fourier Transform. The price for the simplicity of our scheme
is the information rate loss mentioned above: we can now only store $3^{N/2}$ unique ternary sequences on the
medium as opposed to $2^N$ binary sequences - this corresponds to a rate of $\log_2(3)/2\approx 0.79$ bits.
The loss occurs as we ignore the information stored in real parts of $f(n)$'s.
What we gain
however is the ability to store and retrieve information optically using low complexity signal processing.

To modify the described scheme to account for the effects of channel noise we need to replace
step (\ref{eq:trit_recovery_no_noise}) with a maximum-likelihood sequence detector. To derive such a detector,
we need to specify the model of channel noise, which is done in the next subsection.

\subsection{Noise Model}\label{sec:noise}
The experimental characterisation of the optical readout channel is yet to be carried
out. In this paper we restrict ourselves to the simplest additive white Gaussian (AWG) model of
noise in the system. We expect the important sources of noise to be media imperfection, electronics
noise and shot noise. It is a matter for future investigation to verify whether
these can be modelled by AWG noise.

Within the AWG noise model, the received signal is modeled by adding a noise term to the output
of the DFFT: for $n=1,\ldots,N/2$:
\begin{eqnarray}
R_n = \Im(f(n)) + \sigma W_n = \sin(2ks)\sum_{p=0}^{n-1}t_{p} + \sigma W_n
\end{eqnarray}
where $\{W_n\}_{n=1}^{N/2}$ is a  sequence of independent mean-zero Gaussian random variables with unit
variance and $\sigma$ is the standard deviation of the electronics noise.

To define signal-to-noise-ratio (SNR), notice that in the absence of noise, formula (\ref{eq:trit_recovery_no_noise})
is exact. Therefore, it is sensible to define the SNR in terms of the \emph{increments}
of the received sequence $\{R_n\}$ as signals:
\begin{eqnarray}
r_n = R_{n+1}-R_n = sin(2ks)t_n + \sigma (W_n-W_{n-1}).
\end{eqnarray}
SNR measured in decibels is:
\begin{eqnarray}
SNR_{dB} = 10\log_{10}\left(\frac{\sin(2ks)^2}{3\sigma^2}\right)
\end{eqnarray}
It is clear that in order to maximise SNR for a fixed $\sigma$ the indentation
depth $s$ should be chosen so that $sin(2ks)=1$ - the optimal indentation
depth is therefore given by $$s_{opt}=\lambda/8.$$
For a typical wavelength $\lambda\sim 600$ nm, the above formula gives the optimal indentation
depth of $75$ nm. In reality, the indentation depth in probe storage is about $5$ nm. It may be
possible to increase the effective deflection size by incorporating some kind of mechanical amplification
into the cantilever design. In this paper we will simply identify the smallest indentation
depth for which the optical readout is possible without modifying the cantilevers.

\subsection{Experimental Verification of the Diffraction Model}\label{subsec:exper}

\begin{figure}
\begin{centering}
\includegraphics[width=3.3in]{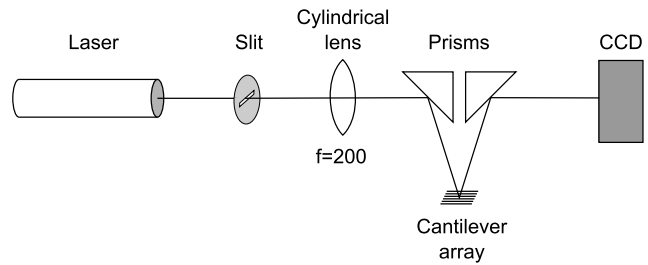}
\par\end{centering}
\caption{Experimental setup of the optical path. A rectangular shaped slit and a
cylindrical lens create a line shaped laser spot on the cantilever array. The array is
positioned in the back focal plane of the lens. The Fraunhofer diffraction pattern is projected on the CCD camera.}
\label{fig:exp_setup}
\end{figure}

To verify the accuracy of the channel model we compare the diffraction
patterns computed within the Fraunhofer theory against experimental
diffraction patterns captured using the setup shown in Figure \ref{fig:exp_setup}.
The array consists of $N = 5$ non-deflected cantilevers of width $w = 13.9~\mu$m whose
centres are separated by $d = 20~\mu$m.
It is illuminated with a laser light source of wavelength $\lambda =$~635~nm and a power of 3~mW.

Although we ultimately foresee the usage of a one dimensional array of very fast photodiodes
to record the diffraction patterns, here we employ a CCD camera (Moticam 2300, Motic, Wetzlar, Germany)
with a two dimensional array of CCD elements (2048 x 1536 pixels) to ease the alignment procedure.
The measurement bandwidth is 111 Hz. The frame rate of the
camera is, however, limited to eight frames per second due to the limited data transfer
rate to the measurement computer.
A low pass filter is applied to the measured intensity pattern to eliminate
spatial noise with frequencies higher than the width of the information band.

\begin{figure}
\begin{centering}
\includegraphics[width=3.3in]{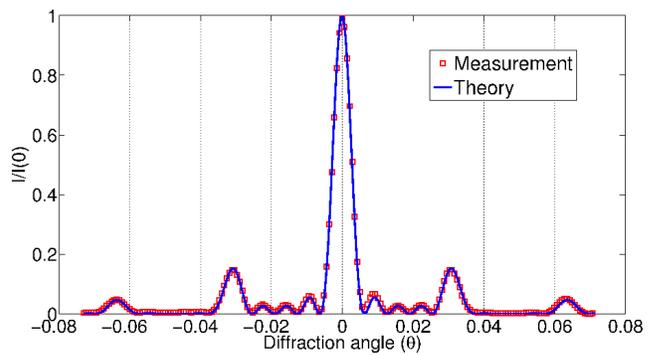}
\par\end{centering}
\caption{Comparison of (\ref{eq:intensity}) with the experimental diffraction
pattern created by the array of $N=5$ undeflected
cantilevers.}
\label{fig:model_verif}
\end{figure}

The experimental and theoretical diffraction curves are shown in Fig. \ref{fig:model_verif}.
An excellent agreement between the theory and the experiment is observed:
the relative error derived using the $L^2$-distance between the theoretical and the experimental curves
is $3.0$ per cent, which is approximately $30.6$ decibel.

\section{Detection Algorithms}\label{sec:detect}

\subsection{Threshold Detector}

In order to derive the maximum-likelihood threshold detector we take the
increment of the DFFT outputs as in (\ref{eq:trit_recovery_no_noise}) and
divide by $\sin(2ks)$ so the received signal $y_n$ for $n=0,\ldots,N/2-1$ can be written:
\begin{eqnarray}
y_n = \frac{R_{n+1}}{\sin(2ks)} - \frac{R_n}{\sin(2ks)} = t_n + \frac{\sqrt{2}\sigma}{\sin(2ks)}W_n\label{eq:recv_thd}
\end{eqnarray}
Then the task of the threshold detector is to examine received signals one-by-one and for each $y_n$ infer the most likely
recorded trit $\hat{t}_n$:
\begin{eqnarray}
\hat{t}_{n} &=& \argmax_{t\in\{-1,0,+1\}}\Pr( t_{n} = t | y_n  )\nonumber
\end{eqnarray}
Applying Bayes' theorem and assuming all trits are equiprobable the maximum-likelihood criterion can be re-written:
\begin{eqnarray}
\hat{t}_{n} &=& \argmax_{t\in\{-1,0,+1\}}\rho(  y_n | t_{n} = t )\label{eq:thd_mlcrit}
\end{eqnarray}
where $\rho(y_n|t)$ is the Gaussian probability density function given by:
\begin{eqnarray}
\rho(y_n | t_{n} = t) = \left(2\pi\sigma_{thd}^2\right)^{-1/2}\exp\left[-\frac{(y_n-t)^2}{2\sigma_{thd}^2}\right]
\end{eqnarray}
where $\sigma_{thd}=\sqrt{2}\sigma/\sin(2ks)$ is the effective standard
deviation of noise in formula (\ref{eq:recv_thd}). Thus the maximum-likelihood
criterion can be reduced to the following \textit{slicing} of the signal:
\begin{eqnarray}
\hat{t}_{n} &=& \left\{\begin{array}{rl}
-1 & \mbox{ if $y_n < -\frac{1}{2}$ } \\
0 & \mbox{ if $-\frac{1}{2} \leq y_n \leq +\frac{1}{2}$ } \\
+1 & \mbox{ if $+\frac{1}{2} < y_n $ }\end{array}\right.
\end{eqnarray}

\subsection{Sequence Detector}

The maximum-likelihood sequence detector examines the whole
sequence of received signals $\vect{Y}=\left(Y_1,\ldots,Y_{N/2}\right)$ where $Y_n$ is given by:
\begin{eqnarray}
Y_n = \frac{R_n}{\sin(2ks)} = T_n(t_{0},\ldots,t_{n-1}) + \frac{\sigma}{\sin(2ks)}W_n\label{eq:recv_vit}
\end{eqnarray}
where $T_n(t_{0},\ldots,t_{n-1})\in \{-n,\ldots,+n\}$ is the sum of the first $n$ central trits:
\begin{eqnarray}
T_n(t_{0},\ldots,t_{n-1}) = \sum_{p=0}^{n-1}t_{p}
\end{eqnarray}
The sequence detector then determines the most likely sequence $\vect{\hat{T}}=(\hat{T}_1,\ldots,\hat{T}_{N/2})$
\begin{eqnarray}
\vect{\hat{T}} = \argmax_{\vect{T}}\Pr( \vect{T} | \vect{Y} )\label{eq:mlvitcrit}
\end{eqnarray}
The above maximum likelihood problem  can be easily placed in a classical context:
by construction the central trits are independent random variables uniformly distributed over the
set $\{-1,0,1\}$. Therefore, the sequence of partial
sums $T_1, T_2, \ldots, T_{N/2}$ is a random walk on the set of integer numbers, \cite{Durret}. Therefore,
(\ref{eq:mlvitcrit}) can be stated as follows: find the most likely trajectory of the random walker originating
at $0$ given the set of noisy observations of the walkers's position at times $1,2,3,\ldots$.

Once the problem ((\ref{eq:mlvitcrit})) is solved, the most likely sequence of trits can be
 easily computed from the sequence $\vect{\hat{T}}$ since for $n=0,\ldots,N/2-1$
\begin{eqnarray}
\hat{t}_{n} = \hat{T}_{n+1} - \hat{T}_n
\end{eqnarray}
where $\hat{T}_0=0$. Again we can invoke Bayes' Theorem and assuming all sequences
$\vect{T}$ are equiprobable the maximum-likelihood criterion (\ref{eq:mlvitcrit}) can be expressed:
\begin{eqnarray}
\vect{\hat{T}} &=& \argmax_{\vect{T}}\rho( \vect{Y} | \vect{T} ) =
\argmax_{\vect{T}}\prod_{n=1}^{N/2}\rho( Y_n | T_n )\label{eq:mlvitcrit2}
\end{eqnarray}
The conditional probability density function $\rho(Y_n|T_n)$ is given by:
\begin{eqnarray}
\rho(Y_n | T_n = T) = \left(2\pi\sigma_{vit}^2\right)^{-1/2}\exp\left[-\frac{(Y_n-T)^2}{2\sigma_{vit}^2}\right]
\end{eqnarray}
Where $\sigma_{vit}=\sigma/\sin(2ks)$ is the standard deviation
of the additive noise appearing in equation (\ref{eq:recv_vit}).
Substituting into (\ref{eq:mlvitcrit2}) we arrive at the following expression for the most-likely sequence $\vect{\hat{T}}$:
\begin{eqnarray}
\vect{\hat{T}} &=& \argmin_{\vect{T}}\sum_{n=1}^{N/2}(Y_n-T_n)^2\label{eq:mlvitcrit3}
\end{eqnarray}
We will now demonstrate that the solution to (\ref{eq:mlvitcrit3}) can be
computed recursively using a graph-based algorithm that closely resembles the
Smith-Waterman algorithm used for genetic sequence matching \cite{Smith1981}.

\begin{figure}
\begin{centering}
\includegraphics[width=2.0in]{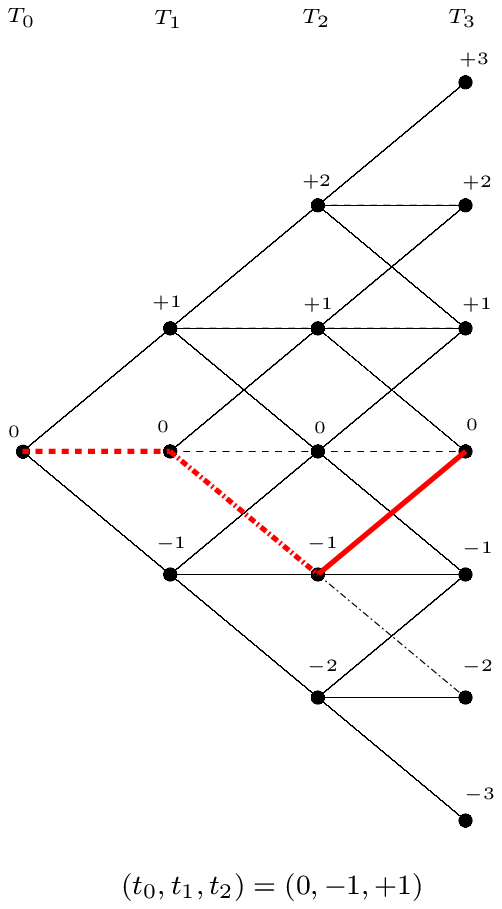}
\par\end{centering}
\caption{A sequence of partial sums of $3$ central trits is drawn on the graph}
\label{fig:graph}
\end{figure}

Consider the graph with $N/2+1$ time slices indexed by $n=0,\ldots,N/2$.
At time slice $n$ we draw $2n+1$ vertices (states) corresponding every
possible value of the variable $T_n$ (the sum of the first $n$ central trits).
Note that the time slice zero consists of only the zero state.
We draw an edge (branch) between the state $T_{n-1}$ and the
state $T_n$ if $T_{n-1}+t = T_{n}$ where $t\in\{-1,0,+1\}$ is a trit. Thus $T_{n}$ is
connected to at most 3 states on the left. If $T_{n}$ is connected to $T_{n-1}$ on the left
we say $T_{n-1}\in L(T_n)$. A path of length $n+1$ is is a sequence of states $(T_0,T_1,\ldots,T_{n})$
such that the states are connected by edges, that is for $m=0,\ldots,n-1$, $T_m\in L(T_{m+1})$.
Thus by construction every sequence $\vect{T}=(T_1,\ldots,T_{N/2})$ can be drawn as a unique
path of length $N/2+1$ through the graph starting from the zero state. A sequence of partial sums of the sequence
of $3$ central trits is drawn on an example graph in Figure \ref{fig:graph}.

\begin{figure}
\begin{centering}
\includegraphics[width=3.5in]{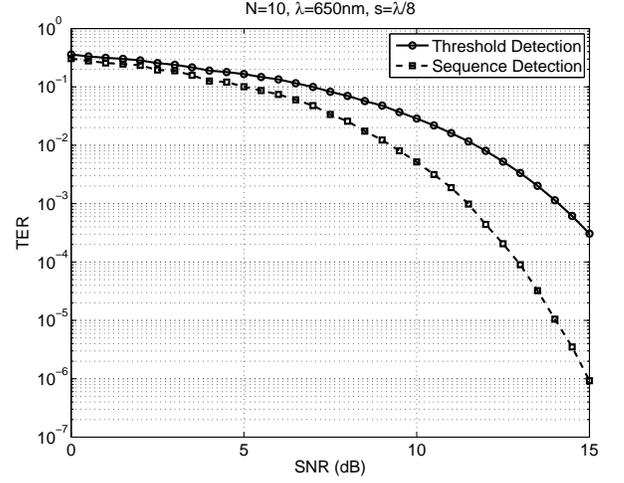}
\par\end{centering}
\caption{TER performance of detection schemes}
\label{fig:detector_ter}
\end{figure}

To each state in the graph we assign a metric $\mu_n(T_n)$ associated with the
probability of passing through state $T_n$:
\begin{eqnarray}
\mu_n(T_n) = (Y_n-T_n)^2
\end{eqnarray}
Next for every path of length $n+1$ through the graph $(T_0,T_1,\ldots,T_n)$
we assign a path metric $\textrm{PM}_n(T_0,T_1,\ldots,T_n)$:
\begin{eqnarray}
\textrm{PM}_n(T_0,T_1,\ldots,T_n) = \sum_{m=1}^n\mu(T_m)
\end{eqnarray}
It can be seen that the maximum likelihood sequence is the path through the
whole graph of the lowest weight:
\begin{eqnarray}
\vect{\hat{T}} = \argmin_{\vect{T}}\textrm{PM}_{N/2}(0,T_1,\ldots,T_{N/2})
\end{eqnarray}
Now we set about constructing the recursion to find this maximally likely path.
Let $P_n(T_n)$ be the set of all paths of length $n+1$ that end in state $T_n$ then we
define the surviving path metric at state $T_n$ to be the minimum weight of all paths that end at state $T_n$:
\begin{eqnarray}
\textrm{SPM}_n(T_n) = \min_{P_n(T_n)}\textrm{PM}_n(T_0,T_1,\ldots,T_n)
\end{eqnarray}
Now the crucial observation is that all paths ending in $T_n$ must have passed
through one of the states $T_{n-1}\in L(T_n)$ so the set $P_n(T_n)$ can be written as the union:
\begin{eqnarray}
P_n(T_n) = \bigcup_{T_{n-1}\in L(T_n)}P_{n-1}(T_{n-1})\label{eq:union}
\end{eqnarray}
The surviving path metric $\textrm{SPM}_n(T_n)$ can be expressed in terms of the
surviving path metrics at the states connected to $T_n$ on the left:
\begin{eqnarray*}
\textrm{SPM}_n(T_n) = \min_{T_{n-1}\in L(T_n)}\left[\textrm{SPM}_{n-1}(T_{n-1})\right] + \mu_n(T_n)
\end{eqnarray*}
Thus the surviving path metrics can be computed for the whole graph recursively.
To find the most likely sequence we then traceback through the graph following the
most likely path. Firstly we find the most likely final state by choosing the state
$T_{N/2}$ with the minimum surviving path metric:
\begin{eqnarray}
\hat{T}_{N/2} = \argmin_{T_{N/2}}\textrm{SPM}_{N/2}(T_{N/2})
\end{eqnarray}
Then we traceback through the graph from the state $\hat{T}_{N/2}$ following the
path of minimum weight. For $n=N/2-1,\ldots,0$:
\begin{eqnarray}
\hat{T}_{n} = \argmin_{T_n\in L(\hat{T}_{n+1})}\textrm{SPM}_n(T_n)
\end{eqnarray}
 To summarise we now describe the detection algorithm in full:
\begin{algorithm}
Maximum-likelihood sequence detector\\
\textbf{Inputs} : $Y_1,\ldots,Y_{N/2}$ \\
\textbf{Outputs} : $t_{N-1,0},\ldots,t_{N/2,N/2-1}$ \\
\begin{algorithmic}[1]
\STATE $\textrm{SPM}_0(0)=0$
\FOR{$n=1$ to $N/2$}
\FOR{$T_n=-n$ to $+n$}
\STATE $\textrm{SPM}_n(T_n) = \min_{T_{n-1}\in L(T_n)}\left( \textrm{SPM}_{n-1}(T_{n-1})\right) + \mu_n(T_n)$
\ENDFOR
\ENDFOR
\STATE $\hat{T}_{N/2} = \argmin_{T_{N/2}}\textrm{SPM}_{N/2}(T_{N/2})$
\FOR{$k=N/2-1$ to $0$}
\STATE $\hat{T}_{n} = \argmin\limits_{T_n\in L(\hat{T}_{n+1})}\textrm{SPM}_n(T_n)$
\STATE $\hat{t}_{n} = \hat{T}_{n+1} - \hat{T}_n $
\ENDFOR
\end{algorithmic}
\end{algorithm}
In Figure \ref{fig:detector_ter} the trit error rate (TER) of the threshold
detector is compared against the TER of the maximum-likelihood sequence detector.
At low error-rates $\textrm{TER}=10^{-4}$ we see an SNR gain of around $2.5dB$.

The complexity of the sequence detection algorithm constructed above is $O(N^2)$ which
should be compared with the exponential complexity of the exhaustive enumeration scheme.
One final remark on the matter of detection is that so far all detection schemes
only make use of the imaginary part of the Fourier coefficients. Information about
the central trits is also contained within the non-linear real part given by (\ref{eq:fnr}).
It may be possible to use the real part of the coefficients as a non-linear parity-check
code if some algebraic or geometrical structure can be identified.

\section{Further considerations for a real system}\label{sec:further}

\subsection{Global positioning errors}

Up to this point we have not accounted for the positioning errors that are present in any
real probe storage device if the information density is higher than $1$ Tb/in$^2$:
to read a series of indentations, the array of probes has
to be moved to the correct sampling position with a nanometer precision, which cannot be done
without incurring positioning errors, see \cite{Harris} for more details.

For an optical system the effect of a global positioning
error is a global reduction in pit depth $s$. Let us assume that the array of $N$
cantilevers at time $i$ suffers a global Gaussian positioning error $J_i\sim\mathcal{N}(0,\sigma_J^2)$
where $\sigma_J$ is the standard deviation of jitter noise. Then the
imaginary part of the Fourier transform output for $n=1,\ldots,N/2$ is:
\begin{eqnarray}
R_n^{(i)} = \sin(2ks_i)\sum_{p=0}^{n-1}t_{p} + \sigma W_n,
\end{eqnarray}
where as before $R_0^{(i)}=0$ and effective pit depth $s_i$ is given by:
\begin{eqnarray}
s_i = \exp\left(-\frac{J_i^2}{w^2}\right)  s.
\end{eqnarray}
The pit depth is reduced by a multiplicative factor given by the simplest probe storage
impulse response \cite{Parnell2009} $p(J)=\exp\left(-J^2/PW^2\right)$ where $PW$ is a parameter related to pulse width.

All of the detection algorithms discussed so far take as input the DFFT output
divided by $\sin(2ks)$, see equations (\ref{eq:recv_thd}) and (\ref{eq:recv_vit}).
Now, in the presence of global positioning errors, the signal degradation due to the loss of
pit depth (jitter) is a hidden random variable.
It is possible however, to estimate the global signal degradation using techniques developed in \cite{Parnell2010}.

Consider the following random variable defined for $n=0,\ldots,N/2-1$ for a fixed time slice $i$:
\begin{eqnarray}
r_n^{(i)} = R_{n+1}^{(i)} - R_n^{(i)} = \sin(2ks_i)t_n + \sigma (W_{n+1}-W_{n})\label{eq:rv}
\end{eqnarray}
We will show that signal degradation due to jitter can be estimated using the empirical average of the square of $r_n^{(i)}$:
\begin{eqnarray}
\overline{RR}_i = \frac{2}{N}\sum_{n=0}^{N/2-1}\left(r_n^{(i)}\right)^2\label{eq:samp_sum}
\end{eqnarray}
As it is easy to see from (\ref{eq:rv}), random variables $\left(r_n^{(i)}\right)^2$ are \emph{not} independent.
However, they are $1$-dependent meaning that subsequences $r^{(i)2}_0~r^{(i)2}_2\ldots r^{(i)2}_p$ and
$r^{(i)2}_{p+k} r^{(i)2}_{p+k+1}\ldots r^{(i)2}_{N/2-1}$ are independent for any $p$ and $k > 1$.
$1$-dependent sequences of random variables (and more generally $M$-dependent sequences for any finite $M$) belong
to the class of the so called $\psi$-mixing correlated sequences for which the strong law of
large numbers can be proved using results of the classical paper \cite{Blum}. (The cited paper established
the strong law of large numbers which is traditionally proved for sequences of independent random variables to the
case of correlated random variables provided the correlations between distant members of the sequence decay at least
exponentially fast.)
Namely, it can be proved that
in the limit $N\rightarrow\infty$ the empirical sum (\ref{eq:samp_sum})
converges almost surely to the corresponding expected value:
\begin{eqnarray}
\overline{RR}_i \overset{a.s}\longrightarrow \mathbb{E}\left[ \left(r_n^{(i)}\right)^2\right]
\end{eqnarray}
This mean can be computed explicitly:
\begin{eqnarray}
\mathbb{E}\left[ \left(R_n^{(i)}\right)^2\right] &=& \frac{2\sin(2ks_i)}{3} + 2\sigma^2
\end{eqnarray}

Thus the signal degradation due to jitter can be estimated by:
\begin{eqnarray}
C_N = \sqrt{\frac{3\left(\overline{RR}_i-2\sigma^2\right)}{2}}\label{eq:jit_est}
\end{eqnarray}
Taking $N$ to infinity this approximation becomes exact:
\begin{eqnarray}
\lim_{N\rightarrow\infty}C_N = \sin(2ks_i)
\end{eqnarray}

Thus it is proposed that the detection schemes discussed in the previous
section can still be used in the presence of global positioning errors by
replacing $\sin(2ks)$ in formulas (\ref{eq:recv_thd}) and (\ref{eq:recv_vit})
with the estimate of $\sin(2ks_i)$: $C_N$. In order for this approximation to
be accurate we need a large number of cantilevers. It may not be practical to
perform optical readback on hundreds or even thousands of cantilevers (since the
Fresnel number which governs the accuracy of Fraunhofer approximation grows with $N$)
but it is possible to structure the probe array
as a large number of rows where each row consists for a relatively small number of
cantilevers (10 for instance). Each row can then be read-back optically and then
the estimate of signal degradation $C_N$ can be computed using the output of the DFFT from all rows.
\begin{figure}
\begin{center}
\includegraphics[width=3.5in]{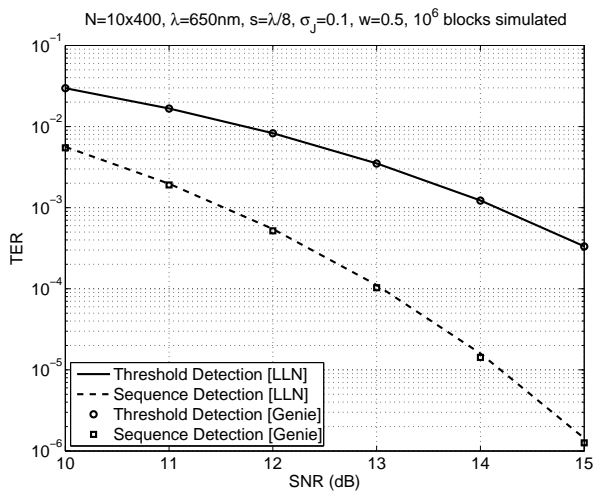}
\caption{Performance of LLN-based detection schemes}
\label{fig:jitter}
\end{center}
\end{figure}

In Figure \ref{fig:jitter} we show results of numerical simulations
which confirm the effectiveness of the proposed detection scheme in the presence
of global positioning errors. We have simulated a cantilever array
with $400$ rows each containing $10$ cantilevers. One million
reads with the array were simulated where at each read
the whole array suffers a global Gaussian positioning error with
standard deviation $\sigma_J=0.1PW$. The two detection schemes
discussed in the previous section have been evaluated
using the estimate (\ref{eq:jit_est}) (LLN detectors) and compared against
the same detection schemes but with perfect knowledge of the signal degradation due to
jitter $\sin(2ks_i)$ (``genie'' detectors). We see that TER is practically the same in both cases.

We conclude with one final remark on the subject of the effect of positioning
errors for optical storage. For the thermo-mechanical read-back channel the loss of
signal strength due to weak jitter ($\sigma_J<<PW$)
can be examined by expanding the impulse response into it's Taylor series around $J=0$:
\begin{eqnarray*}
p(J) = \exp\left(-\frac{J^2}{PW^2}\right) = 1 - \frac{J^2}{PW^2} + O\left(J^3\right)
\end{eqnarray*}
We see that the signal strength decays as $J^2$. For optical read-back the
signal decays (assuming optimal pit depth) can be found by computing the Taylor series:
\begin{eqnarray*}
\sin\left(\frac{\pi}{2}\exp(-J^2/PW^2)\right) = 1 - \left(\frac{\pi^2}{8PW^4}\right)J^4 + O(J^5)
\end{eqnarray*}
Thus we find that for an optical read-back system the jitter noise strength is of the order of $(J/PW)^4$,
which is much smaller than $(J/PW)^2$ - the strength of jitter noise in the thermo-mechanical readout
scheme. The physical reason for a greater resilience of the optical readout scheme to the global jitter
is that to the leading order positioning errors change the phase of the diffraction pattern leaving
the information-carrying intensity invariant.

\subsection{Small pit depth}

In section \ref{sec:noise} it was shown that the optimal pit depth is given by $s=\lambda/8$.
In Table \ref{tab:laser} the optimal pit depth is given for standard laser sources.
The thermo-mechanical write process produces indentations of depth less
than $10$~nm - far smaller than the optimal pit depth even for the latest Blu-ray laser
sources. It is thus necessary to investigate how the performance of the optimal channel
degrades as pit-depth becomes smaller than optimal.

\begin{table}
\begin{center}
  \begin{tabular}{| c | c | c | }
    \hline
    Technology & Wavelength (nm) & Optimal pit depth (nm) \\ \hline
    CD & 780 & 97.50 \\ \hline
    DVD & 650 & 81.25 \\ \hline
    Blu-ray & 405 & 50.63 \\
    \hline
  \end{tabular}
\end{center}
\caption{Optimal pit depths for common laser sources}
\label{tab:laser}
\end{table}

\begin{figure}
\begin{center}
\includegraphics[width=3.5in]{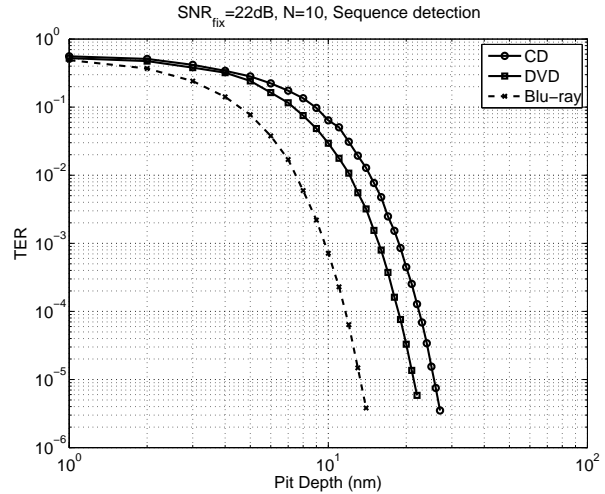}
\caption{Performance degradation due to small pit depth}
\label{fig:pit_depth}
\end{center}
\end{figure}

In order to determine how the channel performance degrades for small sub-optimal
pit depth we must fix the level of noise. For the purpose of this initial
investigation we assume that if pit depth is optimal the signal-to-noise
ratio is fairly high (greater than $20dB$). This is consistent with the pre-high pass filter noise
levels of up to $14$ dB observed in the laboratory \cite{Harris1} although of course in any real system
it is bound to be lower. In Figure \ref{fig:pit_depth} we fix the noise strength so
that for optimal pit depth the SNR is $22dB$ and then plot the TER performance of
the sequence detector as a function of nanometer-scale pit depth for
various standard laser source technologies. We observe that for a Blu-ray
laser source trit error rates of the order $10^{-4}$ can be achieved at pit depths of around $10$~nm.

This initial experiment suggests that it may indeed be possible to
retrieve information at a low error-rate from a probe storage system with
small indentations optically. The next experiment would be to
measure the noise levels for a system with cantilevers
deflected by indentations of a small depth and determine the achievable trit error rate.

\subsection{The effect of modelling errors}
\begin{figure}
\begin{center}
\includegraphics[width=3.5in]{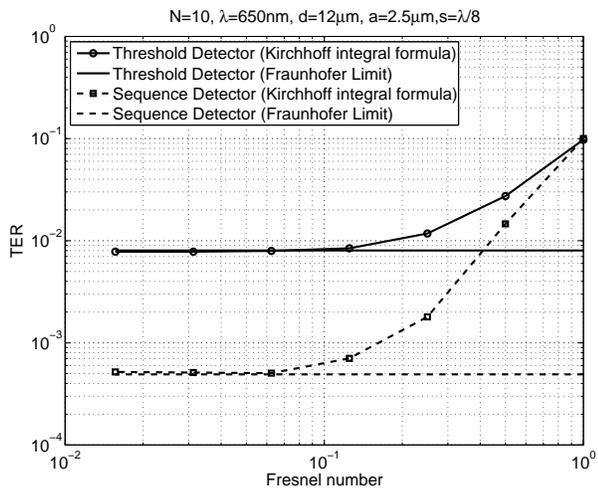}
\caption{Channel performance degrades as the Fresnel number increases.}
\label{fig:model_error}
\end{center}
\end{figure}
So far all signal processing algorithms have been derived
working in the Fraunhofer limit, that is, when the Fresnel number (see Appendix \ref{app:array})  is much less than one:
$$F=\frac{k((N-1)d/2+w/2)^2}{V} \ll 1,$$
where $V$ is the distance between the probe array and the sensor,
see Fig. \ref{fig:geom_strip} for the illustration.
As a concluding experiment we study the degradation of
channel performance if the observation point is moved
closer and closer to the cantilever array so that we gradually move out
of the region of applicability of the Fraunhofer diffraction formula.

In Figure \ref{fig:model_error} we demonstrate how the signal
processing algorithms we have derived degrade with the Fresnel number.
The intensity distribution is
computed by numerically evaluating the Kirchoff diffraction
formula (\ref{eq:kif_array1}) directly. The resulting pattern is
pre-filtered and sent to the DFFT signal
processor as detailed in the section II. Information is extracted from
the output of DFFT using the detection
algorithms derived in the previous section. Recall that all these algorithms
are based on the assumption that the Fraunhofer approximation is valid.

We observe
that if $F < 0.1$ there is virtually no performance
degradation but for $F > 0.1$ the TER increases rapidly:
for $F=1$ the TER of the sequence detector is over two
orders of magnitude higher than the TER within the Fraunhofer limit.
We conclude that in order to use the optical data retrieval techniques
derived in this paper we must ensure the Fresnel number $F < 0.1$.

\section{Conclusions}\label{sec:conclusions}

The paper develops a scheme for storing and retrieving information
in the thermo-mechanical probe storage device using
diffraction patterns created by illuminating the array of actuated probes.
Instead of recovering the information independently by individual probes, we propose
to retrieve information stored in parallel, by analyzing the diffraction patterns
created by groups of $N>>1$ probes. The scheme we developed is slightly reminiscent of
partial response channel equalization used in hard drives in the sense that instead of
trying to avoid diffraction, we detect the most likely information sequence to give rise
to the observed diffraction pattern. The optical readout scheme offers several
advantages, over a more traditional thermo-mechanical method of reading the information off
the probe storage device: firstly, optical sensing is generally faster than thermo-resistive
sensing due to long equilibration times for the latter; secondly, each bit of information
stored in a diffraction pattern is effectively spread over a region of storage medium, thus
making the scheme more resilient to strong instances of media or electronics noise influencing
an individual probe; thirdly, there is evidence that the scheme developed in the paper
offers a better protection of information against global positioning errors - the main
contributor to performance degradation in the nano-scale probe storage.

The principle step which enabled us to construct the low complexity
optical readout scheme is the design of a modulation map between user data and the sequence
of indentations written on the media. The special property of this map is a
linear relation between user information in the
balanced ternary form and the imaginary part of the Fourier coefficients
of the data-dependence diffraction patterns. The key is that the patterns themselves
are generated by the array of probes actuated over the storage medium modulated in accordance
with the user information.

In order to investigate the performance of the proposed optical readout scheme,
a channel model has been derived for the diffraction pattern
obtained when illuminating the cantilever array of a probe
storage device and it is shown that the relative error between
the model and the diffraction patterns gathered experimentally is
less than $3\%$.

It is then shown that by storing ternary information inside the
diffraction patterns it is possible to retrieve this information in the
Fourier domain using low-complexity detection algorithms. In particular a
maximum-likelihood detector has been derived that determines the most likely
sequence of trits by finding the most likely path recursively on a graph.
This detector out-performs the simple threshold detector by $2.5dB$.
The optimal detection algorithm is a version of a two-dimensional Viterbi algorithm
reminiscent of Smith-Waterman sequence matching algorithm used in computational biology.
The low-complexity of the derived detection algorithms can be used to find an optimal
trade-off between the throughput of the detector and its implementation complexity.

The effects of global positioning errors on an optical probe storage system have
been studied and it is shown that the highly parallel nature of probe
storage can be harnessed to compute an accurate estimate of the resulting
global signal degradation (the law of large numbers). The detection algorithms fitted with this estimate
perform at the same error-rates as the
detectors with prior knowledge of positioning errors for a large probe array.
Furthermore, it is shown that the signal degradation that occurs as a result of
positioning errors is less severe for an optical read-back system than the thermo-mechanical equivalent
provided that the amount of probes' deflection is optimal and equal to $\lambda/8$.

Finally, it is found that for levels of noise consistent with our initial
experiments as well experiments with thermo-mechanical probe storage,
is is possible to retrieve information at low error
rates ($TER=10^{-4}$) for sub-optimal pit depths of the order $s$=10~nm.
The degradation in terms of error-rate due to modelling noise has been
quantified and it has been found that as long as the Fresnel number of the
system $F<0.1$, the effects of modelling noise are negligible.

In the present paper we only deal with the problem of storing information
in diffraction patterns generated by a one dimensional probe (sub-)array.
The two-dimensional generalization of the optical readout scheme and the complexity analysis
of the associated signal processing algorithms is an open problem.

We believe that the scheme constructed in this paper can find applications outside the field
of data storage. After all, the problem we are solving is a particular instance of a more
general question: What can be said about the surface which gives rise to a given diffraction
pattern when sensed by an array of probes of known geometry?

The questions investigated in the paper lie on  a borderline between several fields - data storage, optics
and information theory to name just a few. To make the paper as self-contained as possible,
we included several appendices, the aim of which is to derive the Fraunhofer formula for the irregular reflective
grid systematically starting from the two-dimensional
wave equation with the appropriate boundary condition. The derivations are based on the material
presented in \cite{Born} and \cite{Roger}.

\section*{Acknowledgments}
This work is partially supported by the European Research Council within the FP6 collaborative project
Probe based Terabit Memory (ProTeM) No. $2005-IST-5-34719$. The authors would like to thank Leon
Abelmann, Oliver Hambrey, Urs D\"urig and Haris Pozidis for numerous useful discussions.
Martin Siekman and Olti Pjetri are acknowledged for their contribution to the experimental work.

\begin{appendix}

\section{Kirchhoff Integral Theorem}
Let $U$, $U'$ be two solutions to Helmholtz equation in domain $D$ bounded by the closed
curve $S=\partial D$:
\bea
(\Delta+k^2)U(\rv)=0=(\Delta+k^2)U'(\rv), \rv \in D,
\label{heq}
\eea
where $k=\omega/c$, see Fig. \ref{fig:kirkohff}.
According to Green's theorem,
\bea
\int_{D}dV div\Fv=\int_{S}d\sv \cdot \Fv,
\label{gthm}
\eea
where $\Fv$ is an arbitrary smooth vector field, $d\sv=\nv ds$, where $\nv$ is the inward unit
normal vector to the curve $S$, see Fig. \ref{fig:kirkohff}.

Using Green's theorem and Helmholtz equation,
\begin{eqnarray*}
0 &\stackrel{(\ref{heq})}{=}& \int_{D}dV\left\{ U \Delta U'-U'\Delta U   \right\}
\\&=&\int_{D}dV\left\{ div(U\nabla U')-div(U'\nabla U)   \right\}\\
&\stackrel{(\ref{gthm})}{=}&\int_{S}d\sv\cdot \left( U\nabla U' -U' \nabla U\right)
\end{eqnarray*}
We conclude that
\bea
\int_{S}d\sv\cdot \left( U\nabla U' -U' \nabla U\right)=0,
\label{eq:kirch_thm}
\eea
for any pair of solutions to (\ref{heq}) in domain $D$ bounded by a closed curve $S$.
Note that the curve $S$ does not have to be connected.

\section{Kirchhoff formula in two dimensions}

\begin{figure}
\begin{centering}
\includegraphics[scale=0.3]{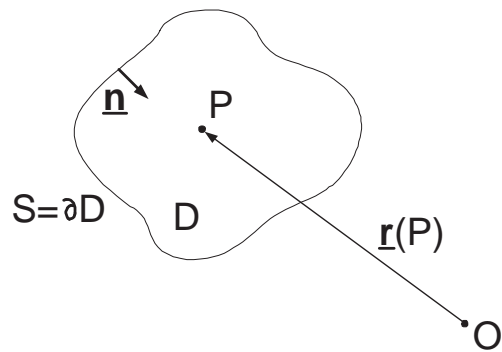}
\par\end{centering}
\caption{Geometric set up for deriving Kirchhoff formula.}
\label{fig:kirkohff}
\end{figure}

Let $G(\rv \mid \rv_0)$ be the Green's function of Helmholtz equation,
\bea
(\Delta+k^2)G(\rv \mid \rv_0)=\delta(\rv-\rv_0),
\label{eq:green_f}
\eea
where $\rv_0$ is the position vector of point $P$, see Fig. \ref{fig:kirkohff}.
It solves homogeneous Helmholtz equation in the region $D\setminus B_{\epsilon}(P)$,
where $B_{\epsilon}(P)$ is a ball of radius $\epsilon$ centered at point $P \in D$.
Applying (\ref{eq:kirch_thm}) to the region $D\setminus B_{\epsilon}(P)$ with $U'(\rv)=G(\rv \mid \rv_0)$
we get
\begin{eqnarray}
&&\int_{S}d\sv\cdot \left( U(\rv)\nabla G(\rv\mid \rv_0) -G(\rv \mid \rv_0) \nabla U(\rv)\right) \label{eq:temp} \\
&=&-\int_{S_{\epsilon}(P)}d\sv\cdot \left( U(\rv)\nabla G(\rv\mid \rv_0) -G(\rv \mid \rv_0) \nabla U(\rv)\right),
\nonumber
\end{eqnarray}
where $S_{\epsilon}(P)=\partial B_{\epsilon}(P)$ is a sphere of radius $\epsilon$ centered at $P$.
We can evaluate the right hand side of the above equation in the limit $\epsilon \rightarrow 0$:
the second term vanishes as $O(\epsilon)$. The first term is equal to
\begin{eqnarray}
-U(\rv_0)\lim_{\epsilon \rightarrow 0}\int_{S_\epsilon(P)}d\sv \nabla G(\rv \mid \rv_{0})\nonumber \\
=-U(\rv_0)\lim_{\epsilon \rightarrow 0}\int_{B_\epsilon(P)}dV \Delta G(\rv \mid \rv_{0}),
\end{eqnarray}
where we applied Green's theorem to convert surface integral to volume integral.
Using (\ref{eq:green_f}) to evaluate $\Delta G(\rv \mid \rv_0)$ and taking the limit
$\epsilon \rightarrow 0$ we find that the right hand side of (\ref{eq:temp}) is equal to $-U(\rv_0)$.
We conclude that
\bea
U(\rv_0)= -\int_{S}d\sv\cdot \left( U(\rv)\nabla G(\rv\mid \rv_0) -G(\rv \mid \rv_0) \nabla U(\rv)\right) \nonumber \\
\eea
In two dimensions,
\bea
G(\rv \mid \rv_0)=\frac{1}{4i} H_0^{(1)}(k \mid \rv-\rv_0 \mid),
\eea
where
\bea
H_0^{(1)}(x)=\frac{2}{\pi i} \int_{1}^{\infty} e^{ixt}(t^2-1)^{-1/2}dt
\eea
is a Hankel function of the first kind.
Therefore we arrive at the following Kirchhoff formula in two dimensions:
\bea
U(\rv_0)&=& \frac{i}{4}\int_{S}d\sv\cdot \bigg( U(\rv)\nabla H_0^{(1)}(k\mid \rv -\rv_0\mid ) \nonumber \\
&-&H_0^{(1)}(k\mid \rv - \rv_0\mid) \nabla U(\rv)\bigg)
\label{eq:kf}
\eea
Formula (\ref{eq:kf}) expresses a solution to Helmholtz equation in terms of its boundary values.
In the context of the theory of electromagnetic waves in two dimensions, $U(\rv)$ describes the spatial distribution
of any component of electromagnetic field strength in a monochromatic wave with wavelength  $2\pi/k$.

\section{Helmholtz-Kirchhoff diffraction theory for a reflective strip}

\begin{figure}
\begin{centering}
\includegraphics[width=2.4in]{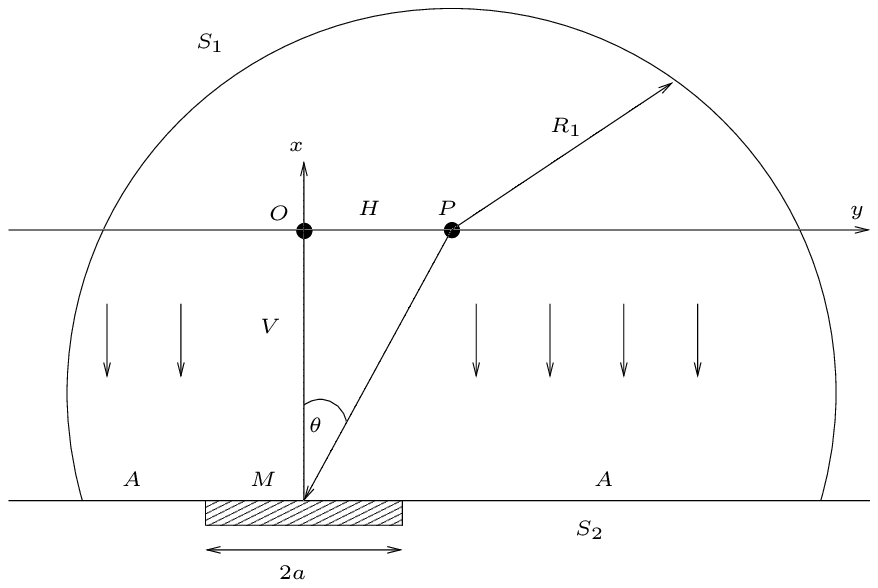}
\par\end{centering}
\caption{Geometry of a reflective strip}
\label{fig:geom_strip}
\end{figure}

We now apply Formula (\ref{eq:kf}) to the two-dimensional geometry given by
Figure \ref{fig:geom_strip} consisting of the observation point $P=(H,0)$
contained with the domain $D$ enclosed by the closed curve $S = S_1 + S_2$
consisting of the arc of a circle of radius $R_1$ around point $P$ ($S_1$) that
intersects with the line $y=-V$ and the line connecting these two points of
intersection ($S_2$). The boundary $S_1=M \cup A$ is split into $M$, the
reflective strip centered at the point $(0,-V)$ with half-width $a=w/2$, and $A$ the aperture.

As $R_1\rightarrow\infty$ the integral on $S_1$ vanishes since the Somerfield
radiation condition is satisfied (only outgoing waves fall on $S_1$) and the field
distribution at $\rv_0(P)$ can be expressed:
\begin{eqnarray}
U(\rv_0) &=& \frac{i}{4}\int\limits_{A+M}d\sv\cdot \bigg( U(\rv)\nabla H(k\mid \rv -\rv_0\mid ) \nonumber \\
&-&H(k\mid \rv - \rv_0\mid) \nabla U(\rv)\bigg)\label{eq:kf_strip1}
\end{eqnarray}
Where $d\sv=-\vect{e}_y$ is the inward normal on $S_2$.
Let $U^{(i)}(\rv)=A_0 e^{i\vect{k}\cdot\rv}$ be the field distribution of the
incident plane wave where $\vect{k}=k(-\vect{e}_y)$. Let $\lambda$ be the
wavelength of the light source (so that $k=2\pi/\lambda$). If the half-width of
the reflective strip is large compared to the wavelength ($ka\gg 1$), then we can use the
Kirchhoff boundary conditions which state:
\begin{enumerate}
\item[(K1)] The field distribution and its derivative are identically zero on $A$: $U|_A = 0$ and $\nabla U |_A = 0$
\item[(K2)] The field distribution and its derivative on $M$ are the same as they would be if $M$ extended over the whole of
$S_2$: $U|_M = A_0e^{i\vect{k}\cdot\rv}$ and $\nabla U |_M = - i \vect{k} U|_M$
\end{enumerate}
If the observation point is far away from the reflective strip so
that $k\mid \rv(x) - \rv_0\mid\gg 1$ then we can use the following
approximations to $H(k\mid \rv - \rv_0\mid)$ and $H'(k\mid \rv - \rv_0\mid)$ \cite{Lebedev}:
\begin{eqnarray}
H(k\mid \rv - \rv_0\mid) &\approx& \sqrt{\frac{2}{\pi k \mid \rv - \rv_0\mid}}e^{i(k\mid \rv - \rv_0\mid - \pi/4)}\\
H'(k\mid \rv - \rv_0\mid) &\approx& i H(k\mid \rv - \rv_0\mid)
\end{eqnarray}
Substituting into (\ref{eq:kf_strip1}) we arrive at the following integral formula for the field distribution at $\rv_0$:
\begin{eqnarray}
U(\rv_0) = \frac{U_0}{4}\int_{-a}^{a}dx\sqrt{\frac{2k}{\pi\mid \rv(x) - \rv_0\mid}}\nonumber\\
\times  \left(1 + \frac{V}{ \mid \rv(x) - \rv_0\mid}\right)e^{ik\mid \rv(x) - \rv_0\mid- i\pi/4}\label{kif_strip_final}
\end{eqnarray}

\section{Intensity of light reflected by an array of probes in Fraunhofer limit}\label{app:array}

We now consider the geometry given in Figure \ref{fig:geom_array}
consisting of an array of $N$ reflecting cantilevers separated by pitch $d$.
The array of cantilevers is positioned above a polymer medium to which the binary
sequence $\vect{b}=(b_0,\ldots,b_{N-1})$ has been recorded by thermo-mechanical
probe storage write process \cite{Pantazi2008}. The presence of a bit $b_n=1$
causes the $n$-th cantilever to be deflected by the indentation depth $s$.
Thus the centre of the $n$-th cantilever is located at position $(H_n,-V_n)$
where  $H_n = nd-(N-1)d/2$ and $V_n=V+b_ns$ for $n=0,\ldots,N-1$ .
The horizontal shift of $(N-1)d/2$ ensures that the cantilevers are positioned
symmetrically with respect to the $y$-axis. As before each cantilever has half-width $a=w/2$.
Illuminating the cantilever array with a plane light source and treating each
cantilever as a reflective strip the field distribution at the observation point $\rv_0=(H,0)$ is given by:
\begin{eqnarray}
U(\rv_0) = \sum_{n=0}^{N-1}\frac{U_n}{4}\int_{H_n-a}^{H_n+a}}dx\sqrt{\frac{2k}{\pi \mid \rv(x) - \rv_0\mid}\nonumber\\
\times  \left(1 + \frac{V}{ \mid \rv(x) - \rv_0\mid}\right)e^{ik\mid \rv(x) - \rv_0\mid- i\pi/4}\label{eq:kif_array1}
\end{eqnarray}
where $U_n=e^{ikV_n}$ is the incident field at the boundary
corresponding to the $n$-th cantilever. This integral is suitable for numerical
evaluation or it is possible to compute it analytically by approximating $|\rv-\rv_0|$ as follows:
\begin{eqnarray}
k|\rv(x)-\rv_0| \approx kR - \frac{kHx}{R} + \frac{kb_nsV}{R}\label{eq:fh_approx}
\end{eqnarray}
Where $R=\sqrt{H^2+V^2}$ is the magnitude of the vector
pointing from the observation point to the centre of the
cantilever array. This approximation is only valid in the
\textit{Fraunhofer limit}, that is, when the Fresnel number is much less than one:
\begin{eqnarray}
F=\frac{k((N-1)d/2+a)^2}{V}\ll 1
\end{eqnarray}
By substituting (\ref{eq:fh_approx}) the integral in
formula (\ref{eq:kif_array1}) can be computed explicitly
resulting in the following equation for the distribution of the amplitude of reflected light
at diffraction angle $\theta = \tan^{-1}(H/V)$:
\begin{eqnarray}
&U&(\theta) = A_0\sqrt{\frac{ka^2}{2\pi R}}\left(1 + \cos(\theta)\right)\frac{\sin(ka\sin(\theta))}{ka\sin(\theta)}e^{i\mu(\theta)}\nonumber\\
&\times& \sum_{n=0}^{N-1}\exp\left[iksb_n(1 + \cos(\theta))-in(kd\sin(\theta))\right]\label{eq:kif_array2}
\end{eqnarray}
where $\mu(\theta) = k(V + R + (N-1)d\sin(\theta)/2)-\pi/4$ is a phase shift
that does not depend on the deflection of the cantilevers.
If $\theta<<1$ (small diffraction angles) we can further simplify (\ref{eq:kif_array2}),
by writing it as a function of parameter $q=k\theta$:
\begin{eqnarray}
U(q) \approx A(\theta) \sqrt{\frac{2ka^2}{\pi R}}\left(\frac{\sin(qa)}{qa}\right)\sum_{n=0}^{N-1}\xi(n)e^{-inqd}
\end{eqnarray}
where $\xi(n) = e^{2iksb_n}$ represents the additional phase gained due to the presence of a deflected cantilever,
$A(\theta)$ is a complex amplitude $\theta$-independent modulus, and $\theta$-dependent phase, which however
does not depend on the state of the cantilever array. As we are only interested in the intensity of reflected
light, the phase of $A(\theta)$ can be set to zero.
\end{appendix}


\begin{thebibliography}{1}
\providecommand{\url}[1]{#1}
\csname url@samestyle\endcsname
\providecommand{\newblock}{\relax}
\providecommand{\bibinfo}[2]{#2}
\providecommand{\BIBentrySTDinterwordspacing}{\spaceskip=0pt\relax}
\providecommand{\BIBentryALTinterwordstretchfactor}{4}
\providecommand{\BIBentryALTinterwordspacing}{\spaceskip=\fontdimen2\font plus
\BIBentryALTinterwordstretchfactor\fontdimen3\font minus
  \fontdimen4\font\relax}
\providecommand{\BIBforeignlanguage}[2]{{%
\expandafter\ifx\csname l@#1\endcsname\relax
\typeout{** WARNING: IEEEtran.bst: No hyphenation pattern has been}%
\typeout{** loaded for the language `#1'. Using the pattern for}%
\typeout{** the default language instead.}%
\else
\language=\csname l@#1\endcsname
\fi
#2}}
\providecommand{\BIBdecl}{\relax}
\BIBdecl

\bibitem{Pantazi2008}
A.~Pantazi, A.~Sebastian, T.~A. Antonakopoulos, P.~B\"achtold, A.~R. Bonaccio,
  J.~Bonan, G.~Cherubini, M.~Despont, R.~A. DiPietro, U.~Drechsler, U.~D\"urig,
  B.~Gotsmann, W.~H\"aberle, C.~Hagleitner, J.~L. Hedrick, D.~Jubin, A.~Knoll,
  M.~A. Lantz, J.~Pentarakis, H.~Pozidis, R.~C. Pratt, H.~E. Rothuizen,
  R.~Stutz, M.~Varsamou, D.~Weismann, and E.~Eleftheriou,
  ``\BIBforeignlanguage{English}{Probe-based ultrahigh-density storage
  technology},'' \emph{\BIBforeignlanguage{English}{IBM J. Res. Dev.}},
  vol.~52, no. 4-5, pp. 493--511, 2008

\bibitem{Manalis1996}
S.~R. Manalis, S.~C. Minne, A.~Atalar, and C.~F. Quate, ``Interdigital
  cantilevers for atomic force microscopy,'' \emph{Appl. Phys. Lett.}, vol.~69,
  no.~25, pp. 3944--3946, 1996.

\bibitem{Yaralioglu1998}
G.~G. Yaralioglu, A.~Atalar, S.~R. Manalis, and C.~F. Quate, ``Analysis and
  design of an interdigital cantilever as a displacement sensor,'' \emph{J.
  Appl. Phys.}, vol.~83, no.~12, pp. 7405--7415, 1998.


\bibitem{Sulchek2001}
T.~Sulchek, R.~J. Grow, G.~G. Yaralioglu, S.~C. Minne, C.~F. Quate, S.~R.
  Manalis, A.~Kiraz, A.~Aydine, and A.~Atalar, ``Parallel atomic force
  microscopy with optical interferometric detection,'' \emph{Appl. Phys.
  Lett.}, vol.~78, no.~12, pp. 1787--1789, 2001

\bibitem{Thundat2000}
T.~Thundat, E.~Finot, Z.~Hu, R.~H. Ritchie, G.~Wu, and A.~Majumdar, ``Chemical
  sensing in fourier space,'' \emph{Appl. Phys. Lett.}, vol.~77, no.~24, pp.
  4061--4063, 2000.

\bibitem{Koelmans2010}
Koelmans, W. W. and Van Honschoten, J. and De Vries, J. and Vettiger,P.
and Abelmann, L. and Elwenspoek, M. C., ``Parallel optical readout of cantilever arrays in dynamic mode'',
Nanotechnol., vol. 21(39), p. 395503, 2010.

\bibitem{Sache2007}
L.~Sache, H.~Kawakatsu, Y.~Emery, H~Bleuler, ``Massively parallel atomic force microscope with
digital holographic readout,'' \emph{Journal of Physics: Conference Series}, no,~61, pp. 668-672, 2007.

\bibitem{Smith1981}
T.~F. Smith, M.~S. Waterman, ``Identification of common molecular subsequences,''
\emph{Journal of Molecular Biology}, vol. ~147, no.~1, pp. 195-197, 1981.

\bibitem{Parnell2010}
O. ~Hambrey, T.~Parnell, O.~Zaboronski, ``Information theory of
massively parallel probe storage channels,'' \emph{preprint arXiv:1102.0540}, 2011.

\bibitem{Marcellino}
Campardo, Giovanni; Tiziani, Federico; Iaculo, Massimo (Eds.), ``Memory Mass Storage'', Chapter 3, Springer 2011.

\bibitem{Born}
Max Born and Emil Wolf, ``Principles of optics'', 7th edition, Cambridge University Press, 1999.

\bibitem{Rivest} Cormen, Thomas H.; Charles E. Leiserson, Ronald L. Rivest, and Clifford Stein
``Introduction to Algorithms'',  2nd edition, Chapter 30,
MIT Press and McGraw-Hill, 2001.

\bibitem{Nyquist} C. E. Shannon, ``Communication in the presence of noise'',
Proc. Institute of Radio Engineers, vol. 37, no. 1, pp. 10–21, Jan. 1949.
Reprint as classic paper in: Proc. IEEE, vol. 86, no. 2, Feb. 1998.

\bibitem{Knuth}
D. E. Knuth, ``The Art of Computer Programming'', Volume {\bf 2}: ``Seminumerical Algorithms''. Third Edition. Reading, Massachusetts: Addison-Wesley, 1997.

\bibitem{Durret}
R. Durret, ``Probability: Theory and Examples'', Third Edition, Brooks/Cole, 2005.

\bibitem{Harris}
A. Sebastian, A. Pantazi, and H. Pozidis,
``Jitter investigation and performance evaluation of a small-scale probe storage device prototype'',
\emph{IEEE Global Communications Conference}, November 2007, pp. 288--293.

\bibitem{Blum}
J. R. Blum, D. L. Hanson, and L. H. Koopmans,
``On the Strong Law of Large Numbers for a Class of
Stochastic Processes'', Z. Wahrscheinlichkeitstheorie vol. 2, pp. 1-11, 1963.

\bibitem{Harris1}
H. Pozidis, et al., `` Demonstration of Thermomechanical Recording at
$641$ Gbit/in$^2$'', \emph{IEEE Transactions on Magnetics}, Vol. 40, no. 4, pp. 2531--2536, 2004.


\bibitem{Parnell2009}
T.~Parnell, H.~Pozidis, O.~Zaboronski,
``Performance evaluation of the probe storage channel,''
\emph{Proceedings of the 28th IEEE Conference on Global Telecommunications }, pp. 6231-6236, 2009.

\bibitem{Roger} G. Scharf, ``From Electrostatics to Optics. A Coincise Electrodynamics Course'', Springer-Verlag, 1994.

\bibitem{Lebedev}
N.N. Lebedev, Richard R. Silverman (Translator), ``Special Functions and Their Applications'', Prentice Hall, 1965.


\end{thebibliography}
\end{document}